\documentclass[letterpaper, 10 pt, conference]{ieeeconf}   
\IEEEoverridecommandlockouts                               
\usepackage{optidef}

\usepackage{amssymb,amsmath,color,subfigure}
\usepackage{graphicx}
\usepackage{xspace,stackrel,hyperref}
\usepackage{wrapfig}
\usepackage{mathtools}
\usepackage{tikz,pgfplots}
\usepackage{etoolbox}
\usepackage{autonum}
\usepackage{dsfont}

\usepackage{makecell}

\usepackage{multirow}
\usepackage{cite}

\usepackage{algorithm, algpseudocode}
\usetikzlibrary{arrows.meta}
\usetikzlibrary{external}

\DeclareMathOperator*{\argmax}{arg\,max}

\usepgfplotslibrary{fillbetween}
\usepgfplotslibrary{groupplots}
\usetikzlibrary{decorations.pathreplacing}

\usepackage{lipsum,adjustbox}

\definecolor{mycolor4}{RGB}{230,97,1}
\definecolor{mycolor2}{RGB}{178,171,210}
\definecolor{mycolor3}{RGB}{253,184,99}
\definecolor{mycolor1}{RGB}{94,60,153}

\makeatletter 
\pretocmd\@bibitem{\color{black}\csname keycolor#1\endcsname}{}{\fail}
\newcommand\citecolor[1]{\@namedef{keycolor#1}{\color{blue}}}
\makeatother

\DeclareMathAlphabet{\mathcal}{OMS}{cmsy}{m}{n}

\def\beq{\begin{equation}}
\def\eeq{\end{equation}}

\newcommand{\mc}{\mathcal}

\newcommand{\Z}{\mathbb{Z}}

\newcommand{\R}{\mathds{R}}

\newcommand{\defineas}{\coloneqq}

\definecolor{mycolor1}{RGB}{230,97,1}
\definecolor{mycolor2}{RGB}{178,171,210}
\definecolor{mycolor3}{RGB}{253,184,99}
\definecolor{mycolor4}{RGB}{94,60,153}
\definecolor{mycolor5}{rgb}{0,0,0}

\tikzset{
  pics/car/.style args={#1}{
     code={
     \begin{scope}[scale=0.15]
      \shade[top color=#1, bottom color=white, shading angle={135}]
        [draw=black,fill=red!20,rounded corners=0.2ex] (1.5,.5) -- ++(0,1) -- ++(1,0.3) --  ++(3,0) -- ++(1,0) -- ++(0,-1.3) -- (1.5,.5) -- cycle;
    \draw[ rounded corners=0.5ex,fill=black!20!blue!20!white]  (2.5,1.8) -- ++(1,0.7) -- ++(1.6,0) -- ++(0.6,-0.7) -- (2.5,1.8);
    \draw[thick]  (4.2,1.8) -- (4.2,2.5);
    \draw[draw=black,fill=gray!50,thick] (2.75,.5) circle (.5);
    \draw[draw=black,fill=gray!50,thick] (5.5,.5) circle (.5);
    \end{scope}
     }
  }
}

\newtheorem{assumption}{Assumption}

\newtheorem{definition}{Definition}

\newtheorem{remark}{Remark}

\newtoggle{full_version}
\toggletrue{full_version}

\title{\LARGE \bf
On Learning-Based Traffic Monitoring With a Swarm of Drones 
}

\author{Marko Maljkovic and Nikolas Geroliminis
\thanks{M.~Maljkovic and N.~Geroliminis are with the School of Architecture, Civil and Environmental Engineering, École Polytechnique Fédérale de Lausanne (EPFL), 1015 Lausanne, Switzerland. {\tt\small \{marko.maljkovic, nikolas.geroliminis\}@epfl.ch}.}%
\thanks{This work was supported by the Swiss National Science Foundation under NCCR Automation Phase 2 project, fonds 565 776.}
\iftoggle{full_version}{}{\thanks{An extended version containing all the proofs is available at \url{https://arxiv.org/abs/???????}}}%
}

\begin{document}

\maketitle
\thispagestyle{empty}
\pagestyle{empty}

\begin{abstract}
Efficient traffic monitoring is crucial for managing urban transportation networks, especially under congested and dynamically changing traffic conditions. Drones offer a scalable and cost-effective alternative to fixed sensor networks; however, deploying fleets of low-cost drones for traffic monitoring poses challenges in adaptability, scalability, and real-time operation. To address these issues, we propose a learning-based framework for decentralized traffic monitoring with drone swarms, targeting the uneven and unpredictable distribution of monitoring needs across urban areas. Our approach introduces a semi-decentralized reinforcement learning model, which trains a single Q-function using the collective experience of the swarm. This model supports full scalability, flexible deployment, and, when hardware allows, the online adaptation of each drone’s action-selection mechanism. We first train and evaluate the model in a synthetic traffic environment, followed by a case study using real traffic data from Shenzhen, China, to validate its performance and demonstrate its potential for real-world applications in complex urban monitoring tasks.

\end{abstract}

\section{Introduction}

Traffic monitoring has become essential for efficient management of urban transportation systems, especially in densely populated areas where congestion is frequent and traffic patterns are highly dynamic~\cite{rs14030620}. Recently, drone technology has emerged as a transformative tool for patrolling missions, offering real-time, high-resolution data for improved situational awareness, all while being able to quickly cover large areas and respond to unexpected changes in traffic demand~\cite{drones7030169}. As cities face growing traffic congestion and the need for effective monitoring intensifies, drone swarms provide a scalable, flexible and affordable alternative to fixed sensor network architectures consisting of loop detectors, infrared sensors or cameras~\cite{9893814, 8377077, 10.1145/3349801.3349805}.

A primary challenge in traffic monitoring with drone swarms is the uneven spatio-temporal distribution of monitoring needs. Downtown areas, for example, experience high traffic demand during rush hours and require frequent monitoring, while city outskirts with lower traffic densities often have a lower priority. In large and unknown urban environments, where full coverage at all times is impractical (and sometimes impossible), traditional approaches to coordinated swarm path planning~\cite{traffica} struggle to address this imbalance effectively, often relying on ad hoc adjustments or manual reprogramming to adapt to changing coverage requirements. Hence, there is a need for an adaptive framework that can account for the dynamics and uncertainties of stochastic traffic patterns, enabling the system to identify and prioritize high-importance areas that require immediate inspection.

Numerous approaches to cooperative multi-agent path planning have been proposed for search, tracking, and patrolling missions~\cite{STACHE2023104288, 111, 6161683, 7365431, 9561014, 9247490, CHEN2021109851, 9811673}. However, these methods often assume perfect sensing, which is impractical for low-cost drones with limited computational capabilities used for traffic data collection. Some are designed exclusively for single-drone setups, while others rely heavily on centralized action planning, creating a single point of failure that could lead to mission abort. In contrast, learning-based approaches offer practical benefits: they enable offline pretraining on existing data, requiring onboard hardware only to perform inference with the trained model. Where hardware capabilities allow, these models can also be refined online during deployment, in both centralized and fully distributed manner. Most works in this area focus on tasks such as target tracking~\cite{10160919, 9623508}, patrolling~\cite{10160923, 222, 9330612}, or collaborative exploration~\cite{9963690, 8929192, LiuICML2021}, and these applications typically fall within the framework of Multi-agent Reinforcement Learning (MARL). 
\begin {figure}
\centering
\begin{adjustbox}{max height=0.35\textwidth, max width=0.48\textwidth}
\begin{tikzpicture}[scale=1.0]
    \def\rows{12}
    \def\cols{15}
    \def\cellsize{0.6} 

    \node (pic1) at (0.0,0.0) {\includegraphics[height=.4\textwidth,width=.5\textwidth]{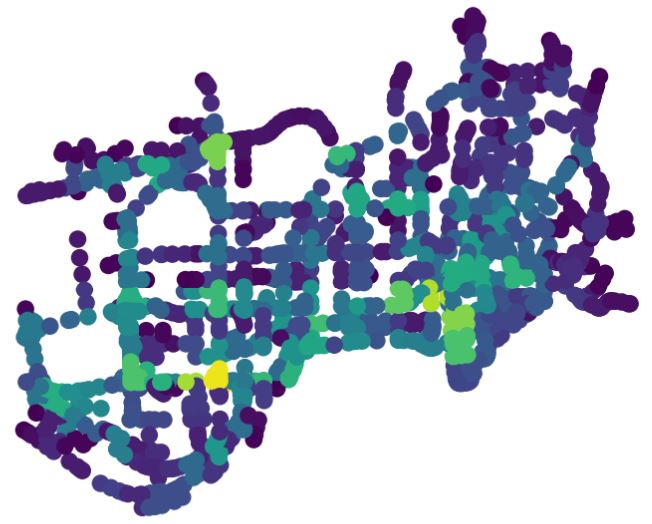}};
    
    \node (pic2) at (-7.5*\cellsize+6.5*\cellsize, 6*\cellsize-3.5*\cellsize) {\includegraphics[height=.03\textwidth,width=.035\textwidth]{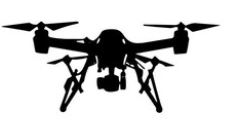}};

    \node (pic3) at (-7.5*\cellsize+3.5*\cellsize, 6*\cellsize-2.5*\cellsize) {\includegraphics[height=.03\textwidth,width=.035\textwidth]
    {figures/drone.JPG}}; 
    \node (pic4) at (-7.5*\cellsize+8.5*\cellsize, 6*\cellsize-8.5*\cellsize) {\includegraphics[height=.03\textwidth,width=.035\textwidth]
    {figures/drone.JPG}}; 
    \node (pic5) at (-7.5*\cellsize+13.5*\cellsize, 6*\cellsize-7.5*\cellsize) {\includegraphics[height=.03\textwidth,width=.035\textwidth]
    {figures/drone.JPG}};

    \foreach \x in {0,...,\cols} {
        \foreach \y in {0,...,\rows} {
            \draw[line width=0.025pt, draw=black!30] (\x*\cellsize-7.5*\cellsize, 0-6*\cellsize) -- (\x*\cellsize-7.5*\cellsize, \rows*\cellsize-6*\cellsize);
            \draw[line width=0.025pt, draw=black!40] (0-7.5*\cellsize, \y*\cellsize-6*\cellsize) -- (\cols*\cellsize-7.5*\cellsize, \y*\cellsize-6*\cellsize);
        }
    }
    
    \draw[-{Triangle[length=3pt, width=2.5pt]},line width=0.5pt](-7.5*\cellsize-0.25, 6*\cellsize+0.25)--(-7.5*\cellsize-0.25+8*\cellsize, 6*\cellsize+0.25);
    \node[ scale=0.8, black]() at (-7.5*\cellsize-0.25+8*\cellsize, 6*\cellsize+0.25+0.25)  {$y$};
    
    \draw[-{Triangle[length=3pt, width=2.5pt]},line width=0.5pt](-7.5*\cellsize-0.25, 6*\cellsize+0.25)--(-7.5*\cellsize-0.25, 6*\cellsize+0.25-5*\cellsize);
    \node[ scale=0.8, black]() at (-7.5*\cellsize-0.25-0.25, 6*\cellsize+0.25-5*\cellsize) {$x$};

    \draw[dotted, line width=0.7pt](-7.5*\cellsize+6.5*\cellsize, 6*\cellsize-3.5*\cellsize)--(-7.5*\cellsize+6.5*\cellsize, 6*\cellsize+0.25);  
    \node[ scale=0.8, black]() at (-7.5*\cellsize-0.25+8*\cellsize-\cellsize, 6*\cellsize+0.25+0.25) {$z_{3,6}^y$};
    
    \draw[dotted, line width=0.7pt](-7.5*\cellsize+6.5*\cellsize, 6*\cellsize-3.5*\cellsize)--(-7.5*\cellsize-0.25, 6*\cellsize-3.5*\cellsize); 
    \node[ scale=0.8, black]() at (-7.5*\cellsize-0.25-0.25, 6*\cellsize+0.25-5*\cellsize+\cellsize) {$z_{3,6}^x$};

    \node[scale=0.6, black]() at (-7*\cellsize, 5.5*\cellsize){$(0,0)$};  
    \node[scale=0.6, black]() at (-6*\cellsize, 5.5*\cellsize){$(0,1)$}; 
    \node[scale=0.6, black]() at (-7*\cellsize, 4.5*\cellsize){$(1,0)$};

    \node[ scale=1.2, black]() at (-7.5*\cellsize-0.25+8*\cellsize, -6*\cellsize+0.25-0.5)  {Area width $W$};
    \node[ scale=1.2, black, rotate=90]() at (7.5*\cellsize+0.3, 0.0)  {Area height $H$};
    
\end{tikzpicture}
\end{adjustbox}
    \caption{Schematic representation of the city of Shenzhen divided into a 12x15 monitoring grid and covered by a swarm consisting of four drones. Nodes of the traffic network representing the city area are color-coded to illustrate the traffic demand at a particular time instance. Drone $d\in\mc D$, located in the cell $(3,6)$, has the corresponding $(x,y)$ world coordinates given by $\mathbf{p}_d=[z_{3,6}^x,z_{3,6}^y]^T$.} 
\label{fig:setup}
\end{figure}

To the best of our knowledge, there exists no universal end-to-end multi-agent monitoring solution, agnostic to the type of traffic phenomena that is being observed in a changing urban environment. Our approach is similar to~\cite{Thaker2023jul, 9330612, 10644645} in a sense that we also consider a grid partitioning of the covered area, but propose a semi-decentralized Reinforcement Learning (RL) swarm control method. Unlike existing approaches that struggle with the curse of dimensionality or become entirely unscalable when jointly training all drones' actions, we design a traffic state representation that enables learning a single Q-function using the collective experience of the entire swarm. This Q-function can be directly deployed on each drone, allowing for individual action-selection process and potential online refinement if permitted by the available hardware. The model is initially trained and compared against traditional methods in an artificial environment. However, the model parameters are then directly transferred to a simulated case study based on the real traffic demand in the city of Shenzhen, showcasing the potential for sim-to-real transfer when tackling more complex monitoring tasks. 

The paper is organized as follows: in the following section, we formally introduce the traffic monitoring problem, along with the models for the drone and the swarm. In Section~\ref{sec:learn}, we present the technical details related to the design and learning process of the Q-function. Finally, in Sections~\ref{sec:example} and~\ref{sec:conclusion}, we analyze the performance of the proposed model, compare it with some traditional patrolling mechanisms, and outline potential ideas for future research.   

\textit{Notation:}  Let $\R_{(+)}$ and $\Z_{(+)}$ denote the sets of (non-negative) real and integer numbers. For any $T\in\Z_+$, we let $\Z_T=\{0,1,2,...,T-1\}$.

\section{Problem Formulation}\label{sec:model}

We consider the problem of monitoring a rectangular dynamic urban map of size $H\times W$ with a swarm of $N$ drones, denoted by $\mc D$, over a time horizon $T\in\Z_+$. The objective of the swarm is to adapt its patrolling behavior to account for dynamically changing spatio-temporal distribution of traffic congestion in order to optimize partial map coverage through time. Each drone can act as a mobile sensor, gathering data on traffic parameters such as flow, density, road occupancy, etc., enabling the fleet to detect congestion in different areas of the map at specific times. However, the coverage priority varies across the map and through time. For instance, the city outskirts typically demand less monitoring than downtown areas during rush hours due to lower traffic volumes and generally slower dynamics. Therefore, based on the observed temporal evolution of the state of traffic, the fleet should assess the relevance of each drone's surrounding and decide how to rebalance.
\subsection{Drone model}

Formally speaking, let $\mathbf{p}_{d}^k\in[0,H]\times[0,W]$ denote the position of drone $d\in\mc D$ at time $k\in\Z_T$. Moreover, let the map be divided into $\mathbf{n}_x\times\mathbf{n}_y$ cells according to Definition~\ref{def:1} and the movement of drones be constrained by Assumption~\ref{ass:1}. A schematic representation of the setup is shown in Figure~\ref{fig:setup}.
\begin{definition}\label{def:1}
A grid partitioning of a rectangular urban map of size $H\times W$ into $\mathbf{n}_x\times\mathbf{n}_y$ cells is defined by a set $\mc M\subset\Z_+^2$ whose elements $z_{i,j}\in\Z^2_+$ represent the centers of grid cells $(i,j)$, i.e.,
$$\mc M\defineas\bigl\{z_{i,j}\in\Z^2_+\mid z^T_{i,j}=[z_{i,j}^x, z_{i,j}^y]\land i\in\Z_{\mathbf{n}_x}, j\in\Z_{\mathbf{n}_y}\bigr\}$$
where vector components $z^x_{i,j}$ and $z^y_{i,j}$ are given by
\begin{equation}\label{eq:1}
    z_{i,j}^x\defineas\frac{(2i+1)H}{2\mathbf{n}_x}\:\land\:z_{i,j}^y\defineas\frac{(2j+1)W}{2\mathbf{n}_y}.
\end{equation}
\end{definition} 
\medskip
\begin{assumption}\label{ass:1}
    For every drone $d\in\mc D$ and time step $k\in\Z_T$, it holds that $\mathbf{p}_{d}^{k}\in\mc M$, i.e., drones are always located in one of the grid cell centers. 
\end{assumption}
\medskip

Let $\mc O$ denote a predefined subset of cells corresponding to fixed obstacles in the map, i.e., fields that cannot be visited: $$\mc O\subset\{(i,j)\in\Z_+^2\mid i\in\Z_{\mathbf{n}_x}\:\land\: j\in\Z_{\mathbf{n}_y}\}.$$ Consequently, let $\mc U_{d}(\mathbf{p}_{d}^{k})$ denote the set of feasible actions of drone $d$ at position $\mathbf{p}_{d}^{k}$, i.e.,
\small
$$\mc U_{d}^{k}(\mathbf{p}_{d}^{k})\subseteq\biggl\{\left[\begin{array}{c}
     \delta_x  \\
     0 
\end{array}\right], \left[\begin{array}{c}
     -\delta_x  \\
      0
\end{array}\right], \left[\begin{array}{c}
     0  \\
     \delta_y 
\end{array}\right], \left[\begin{array}{c}
     0  \\
     -\delta_y
\end{array}\right], \left[\begin{array}{c}
     0  \\
     0 
\end{array}\right]\biggr\},$$
\normalsize
where $\delta_x=\frac{H}{\mathbf{n}_x}$, $\delta_y=\frac{W}{\mathbf{n}_y}$, and such that no $\mathbf{u}_{d}^{k}\in\mc U_{d}^{k}(\mathbf{p}_{d}^{k})$ moves the drone outside the map boundaries or causes it to collide with an obstacle from $\mc O$. As a result, after taking $\mathbf{u}_{d}^{k}\in\mc U_{d}^{k}(\mathbf{p}_{d}^{k})$, the drone's position is simply updated as $$\mathbf{p}_{d}^{k+1}=\mathbf{p}_{d}^{k}+\mathbf{u}_{d}^{k}.$$  
\subsection{Observation model}\label{subsec:obsmod}

Drones on the map do not have access to the full state of the traffic environment. Instead, they have to move around, collect local observations, and discover the dynamics of the traffic phenomena that is being observed. Formally speaking, at every $k\in\Z_T$, drone $d$ located at $\mathbf{p}_{d}^{k}$ observes a relevant traffic variable $z(\mathbf{p}_{d}^{k})\in\R_+$ at its location (e.g., traffic flow, traffic density, road occupancy, etc.), as well as in its adjacent cells. If $\mathbf{p}_d^k$ is represented as $\mathbf{p}_d^k=[z_{i,j}^x, z_{i,j}^y]^T$ for some $z_{i,j}^x$ and $z_{i,j}^y$ given by~\eqref{eq:1}, then drone $d$ observes a set of values $\mc Z_d^k\defineas\{z(\mathbf{p})\in\R_+\mid \mathbf{p}\in\operatorname{Adj}(\mathbf{p}_{d}^{k})\}$ such that $$\operatorname{Adj}(\mathbf{p}_{d}^{k})\defineas\{z_{i,j}\in\mc M\mid (i,j)\in\mc A\},$$ 
$$\mc A\defineas\{(i+1,j),(i-1,j),(i,j-1),(i,j+1),(i,j)\}.$$
The observation model of this form implicitly accounts for the drone's \textit{field of view}. In other words, assuming that the drone can observe the traffic conditions in the adjacent cells implies that the drone's field of view extends beyond just the cell it currently occupies. Clearly, the model can be naturally adjusted to consider a broader field of view or restricted to only the current cell where the drone is located 

Regardless of the type of observation, we assume that the upper and lower bounds $\overline{z},\underline{z}>0$ constrain $z(\mathbf{p}_{d}^{k})$ according to the nature of the corresponding traffic variable, i.e., for all $\mathbf{p}\in\mc M$, it holds that $\underline{z}\leq z(\mathbf{p})\leq \overline{z}$. Using the observed values from different cells, the fleet should interpret the traffic data as an importance score that indicates the urgency of returning a drone to that area. For instance, the drones could measure the number of vehicles in a grid cell. The higher the number of vehicles, the more likely that the area is experiencing high traffic demand and is therefore at higher risk of congestion formation.

From a practical standpoint, advancements in computer vision can nowadays provide accurate estimation of vehicle trajectories that can allow for an accurate calculation of different traffic variables, based on the need of the specific application (i.e., monitoring congestion~\cite{10422673}, emissions~\cite{ESPADALERCLAPES2023103822}, parking~\cite{10534095} etc.). Therefore, there is an inherent need for a unified framework that is independent of the specific traffic variable being monitored and can be also deployed on low-cost monitoring drones. With this in mind, we introduce the \textit{temporal importance value} that maps any type of observed traffic variable into a real value in the interval $[0,1]$.
\begin{definition}\label{def:2}
    For a grid partitioning $\mc M\subset\Z_+^2$ of size $\mathbf{n}_x\times\mathbf{n}_y$ given by Definition~\ref{def:1}, and a set of obstacles $\mc O\subset\mc M$, the temporal importance value at time $k\in\Z_T$, at location $\mathbf{p}\in\mc M$, is given by
\begin{equation}
    \mathbf{T}^k(\mathbf{p})\defineas 
    \begin{cases}
        \frac{z(\mathbf{p})-\underline{z}}{\overline{z}-\underline{z}}, & \text{if } \mathbf{p}\in\mc M\setminus\mc O\\
        0, & \text{if } \mathbf{p}\in\mc O
    \end{cases}.
\end{equation}
\end{definition}
\medskip
By transforming any real-world traffic state observation into a unique importance score, where a higher value indicates a greater level of coverage urgency, it becomes possible to deploy heterogeneous fleets for monitoring. For a drone $d\in\mc D$, we will denote the set of transformed observation values at time $k\in\Z_T$ as $\overline{\mc Z}_d^k$.
\subsection{Drone swarm model}

A key advantage of deploying a swarm of drones for traffic monitoring is the potential to achieve efficient extensive coverage through coordinated planning. However, in large areas, traditional patrolling methods often rely on fixed sweep patterns that overlook the need for flexible, uneven coverage. In that sense, we are interested in observation-driven path planning for each drone in the swarm, allowing them to prioritize regions based on real-time information. To facilitate this, we assume a central coordinator facilitates the sharing of observed information, while each drone selects its own action independently. If control were fully centralized, the fleet's action space would grow exponentially with the number of drones $N$, making coordination increasingly computationally expensive. Moreover, a fully central planner would represent a potential single point of failure in the system that would require grounding the entire fleet if disrupted. Thus, we aim to design a distributed monitoring system that leverages information exchange between drones when possible, yet also supports fully decentralized decision-making in the event of coordination interruption.  

To optimize patrolling behavior over a grid $\mc M$, we assume the drones have access to a shared map representing each cell's \textit{idleness}. Specifically, the fleet should prioritize high-importance zones that have not been observed for extended periods of time over other high-importance zones that were recently visited. Thus, each element of the idleness map should reflect the time elapsed since the last drone visit, with higher idleness values indicating a greater urgency to revisit that area. Formally, we introduce the notion of an \textit{idleness map} as in Definition~\ref{def:3}.
\begin{definition}\label{def:3}
    For a grid partitioning $\mc M\subset\Z_+^2$ of size $\mathbf{n}_x\times\mathbf{n}_y$ given by Definition~\ref{def:1}, the idleness map at time $k\in\Z_T$ is given by a matrix $\mathbf{I}^k\in[0,1]^{\mathbf{n}_x\times\mathbf{n}_y}$, whose $(i,j)$-th element represents the idleness of the corresponding $(i,j)$-th grid cell $\mathbf{p}\in\mc M$ and is denoted by $\mathbf{I}^k(\mathbf{p})$.
\end{definition}
\medskip

To track recently visited cells, the idleness map needs to be constantly updated. Let $\mc P_k\defineas\{\mathbf{p}_{d}^{k}\}_{d\in\mc D}$ denote the set of all drones' positions at time $k$. Then, after the fleet moves to $\mc P_{k+1}$, the idleness map values are updated as:
\small\begin{equation}
    \mathbf{I}^{k+1}(\mathbf{p})=\begin{cases}
        \eta\mathbf{I}^k(\mathbf{p}), & \text{if }\mathbf{p}\in\mc P_{k+1} \\
        \min{(1;\:\mathbf{I}^k(\mathbf{p})+\delta)}, & \text{if }\mathbf{p}\in\mc M\setminus(\mc P_{k+1}\cup\mc O)\\
        0, & \text{if }\mathbf{p}\in\mc O
    \end{cases}.
\end{equation}\normalsize
with $\eta,\delta\in(0,1)$ being the forgetting and recovery factors. Clearly, a smaller $\eta$ value means that a particular area will take longer to reach a state requiring an urgent revisit, while a higher $\delta$ value accelerates this recovery process. 

In essence, to optimize its patrolling behavior with respect to real-time, uneven coverage requirements, the swarm needs to balance between prioritizing a greedy collection of temporal importance values that have not recently been collected $\tilde{\mathbf{T}}^k(\mathbf{p})=\mathbf{I}^{k}(\mathbf{p})\mathbf{T}^{k}(\mathbf{p})$, and a greedy collection of idleness  $\mathbf{I}^k(\mathbf{p})$ with its drones $d\in\mc D$. If we let $$\mathbf{u}_{k}\in\mathbf{U}(\mc P_k)\defineas\bigtimes_{d\in\mc D}\mc U_{d}^{k}(\mathbf{p}_{d}^{k})$$ denote a joint strategy of all drones, then for an action that transitions the fleet from cells $\mc P_k$ to cells $\mc P_{k+1}$, a weighted patrolling score $\mathbf{R}(\mc P_k, \mathbf{u}_k,\mc P_{k+1})\in\R$ in line with Definition~\ref{def:4} naturally imposes.
\begin{definition}\label{def:4}
    For a grid partitioning $\mc M\subset\Z_+^2$ of size $\mathbf{n}_x\times\mathbf{n}_y$ given by Definition~\ref{def:1}, a temporal importance map determined by $\mathbf{T}^k(\mathbf{p})\in\R$ as in Definition~\ref{def:2}, and the idleness map $\mathbf{I}^k\in[0,1]^{\mathbf{n}_x\times\mathbf{n}_y}$ given by Definition~\ref{def:3}, the weighted patrolling score $\mathbf{R}(\mc P_k, \mathbf{u}_k,\mc P_{k+1})\in\R$ of a feasible action $\mathbf{u}_k\in\mathbf{U}(\mc P_k)$ that transitions the fleet from cells $\mc P_k$ to cells $\mc P_{k+1}$ is given by
\small\begin{equation}\label{eq:r}
    \mathbf{R}(\mc P_k, \mathbf{u}_k,\mc P_{k+1})\defineas \sum_{d\in\mc D}\alpha_{\textbf{T}}\overline{\mathbf{T}}_d(\mathbf{p}_{d}^{k+1}, \mathbf{p}_{d}^k)+\alpha_{\textbf{I}}\overline{\mathbf{I}}_d(\mathbf{p}_{d}^{k+1}, \mathbf{p}_{d}^k),
\end{equation}\normalsize
where the individual terms are described by
\small\begin{equation}
   \overline{\mathbf{T}}_d(\mathbf{p}_{d}^{k+1}, \mathbf{p}_{d}^k)\defineas\sum_{\mathbf{p}\in\text{Adj}(\mathbf{p}_{d}^{k+1})}\tilde{\mathbf{T}}^{k+1}(\mathbf{p})-\sum_{\mathbf{p}\in\text{Adj}(\mathbf{p}_{d}^{k})}\tilde{\mathbf{T}}^k(\mathbf{p})
\end{equation}\normalsize
\small\begin{equation}
   \overline{\mathbf{I}}_d(\mathbf{p}_{d}^{k+1}, \mathbf{p}_{d}^k)\defineas\sum_{\mathbf{p}\in\text{Adj}(\mathbf{p}_{d}^{k+1})}\mathbf{I}^{k+1}(\mathbf{p})-\sum_{\mathbf{p}\in\text{Adj}(\mathbf{p}_{d}^{k})}\mathbf{I}^k(\mathbf{p})
\end{equation}\normalsize
and $\alpha_{\textbf{T}},\alpha_{\textbf{I}}>0$ depict the swarm's behavioral preference.
\end{definition}
\medskip
In particular, the term $\overline{\mathbf{T}}_d(\mathbf{p}_{d}^{k+1}, \mathbf{p}_{d}^k)$ rewards actions that direct drones toward areas of high temporal importance. Conversely, $\overline{\mathbf{I}}_d(\mathbf{p}_{d}^{k+1}, \mathbf{p}_{d}^k)$ incentivizes movement toward cells that have gone unobserved for extended periods, whereas the ratio $\alpha_{\textbf{T}}/\alpha_{\textbf{I}}$ determines which type of behavior is prioritized.
\begin{remark}
    The definition of $\overline{\mathbf{T}}_d(\mathbf{p}_{d}^{k+1}, \mathbf{p}_{d}^k)$ can easily be adjusted to account for different fields of view in line with the observation model adopted in Section~\ref{subsec:obsmod}.
\end{remark}
\medskip

Governed by the intuitive, instantaneous `action-score' measure of patrolling performance in an unknown dynamic traffic environment, we turn to learning-based methods and propose a semi-decentralized Reinforcement learning (RL) framework for traffic monitoring with a swarm of drones. The proposed approach is decentralized in a sense that we do not train one agent to choose the action of all drones simultaneously as typically done in the literature. However, we do propose an additional centralized coordination scheme to help avoid undesired scenarios. To evaluate and compare the effectiveness of various action-selection mechanisms, in addition to reporting the cumulative patrolling score over the horizon $T$, i.e., $\mathbf{R}^T=\sum_{k\in\Z_T}\mathbf{R}(\mc P_k, \mathbf{u}_k,\mc P_{k+1})$, we also formally introduce the \textit{coverage score}.
\begin{definition}\label{def:5}
    At time $k\in\Z_T$, for a grid partitioning $\mc M\subset\Z_+^2$ of size $\mathbf{n}_x\times\mathbf{n}_y$ given by Definition~\ref{def:1}, a temporal importance map determined by $\mathbf{T}^k(\mathbf{p})\in\R$ as in Definition~\ref{def:2}, and the idleness map $\mathbf{I}^k\in[0,1]^{\mathbf{n}_x\times\mathbf{n}_y}$ given by Definition~\ref{def:3}, the coverage score $\mathbf{C}^k\in\R$ is given by
    \begin{equation}
        \mathbf{C}^k=1-\frac{\sum_{\mathbf{p}\in\mc M\setminus\mc O}\mathbf{I}^k(\mathbf{p})}{\mathbf{n}_x\mathbf{n}_y-|\mc O|}.
    \end{equation}
\end{definition}
\medskip
A higher coverage score indicates better monitoring of the whole map. Indeed, higher values of $\mathbf{C}^k$ correspond to a lower average idleness value of non-obstacle cells, which suggests that, on average, each cell is more often visited.
\section{Learning based traffic monitoring}\label{sec:learn}

At a particular time step $k\in\Z_T$, the swarm needs to decide on a joint feasible action $\mathbf{u}_k\in\mathbf{U}(\mc P_k)$  in a coordinated or fully distributed manners based on available information. As the number of possible actions of each drone is finite, we propose a distributed approach based on the concept of Q-learning~\cite{qlr}. Specifically, from the available data comprising the locations of all drones in the fleet $\mc P_k$, the local temporal importance observations in the drones' field of view $\bigcup_{d\in\mc D}\overline{\mc Z}_d^k$, and the current idleness matrix $\mathbf{I}^k$, each drone $d\in\mc D$ should interpret the traffic environment as a personalized state vector $\mathbf{s}_d^k\in\R^m$, where $m$ is the state vector dimension, and choose an action in accordance with the optimal policy $\pi_d:\mathbf{s}_d^k\rightarrow\mathbf{u}_d^k$. Given the estimated states, the optimal action-selection policy should evaluate the `quality' of all feasible state-action pairs $(\mathbf{s}_d^k,\mathbf{u})$ and opt for $\mathbf{u}$ that maximizes the Q-function $Q_d:(\mathbf{s}_d^k, \mathbf{u})\rightarrow\R$. 

In the context of traffic monitoring, Q-function calculates drone $d$'s total $\gamma$-discounted patrolling score of choosing action $\mathbf{u}_d^k$ when in state $\mathbf{s}_d^k$, while also considering future developments. In our case, the following Bellman equation has to hold when following the optimal policy
\small\begin{equation}
    Q_d(\mathbf{s}_d^k,\mathbf{u}_d^k)=\mathbf{r}_d^k(\mathbf{s}_d^k,\mathbf{u}_d^k,\mathbf{s}_d^{k+1})+\gamma\max_{\mathbf{u}'\in\mc U_d^{k+1}(\mathbf{p}_{d}^{k+1})}Q_d(\mathbf{s}_d^{k+1},\mathbf{u}'),
\end{equation}\normalsize
where $\mathbf{r}_d^k(\mathbf{s}_d^k,\mathbf{u}_d^k,\mathbf{s}_d^{k+1})\in\R$ denotes the instantaneous reward obtained from taking the action $\mathbf{u}_d^k$.

Since all drones choose actions from the same set of all possible actions, and the inherent reasoning about the quality of the state-action pairs should be the same as well, we adopt a unique parametrized Q-function governing the decision making of all drones, i.e., for all $d\in\mc D$, it holds that $Q_d(\mathbf{s},\mathbf{u})=Q(\mathbf{s},\mathbf{u};\theta)$, where $\theta\in\R^{N_{\theta}}$ is the vector of learnable parameters. In such a setup, if two drones find themselves in the same grid cell at some point, they will continue to follow same paths since their observations, and hence the outputs of their Q-functions, would be the same. This can be alleviated if the centralized coordination is enabled. Namely, if the estimate of the Q-function parameters at time $k\in\Z_T$ is given by $\theta^k$, then the optimal joint action can be determined by solving
    \begin{maxi}
        {\mathbf{u}_k\in\mathbf{U}(\mc P_k)}{\sum_{d\in\mc D}Q(\mathbf{s}_d^k, \mathbf{u}_d^k; \theta^k) \label{maxi:op2}}
        {}{}
        \addConstraint{\mathbf{p}_d^{k+1}=\mathbf{p}_d^k+\mathbf{u}_d^k,\:\forall d\in\mc D}{}{}
        \addConstraint{\mathbf{p}_{d_1}^{k+1}\neq\mathbf{p}_{d_2}^{k+1},\:\forall d_1,d_2\in\mc D, d_1\neq d_2}{}{}
    \end{maxi}
Alternatively, if centralized coordination is disrupted, each drone’s action $\mathbf{u}_k^d$ can be determined using an $\varepsilon$-parametrized stochastic policy to help mitigate potential overlap of drones:
\begin{equation}
    \mathbf{u}_k^d=\begin{cases}
        \displaystyle\argmax_{\mathbf{u}\in\mc U_d^k(\mathbf{p}_d^k)} Q(\mathbf{s}_d^k, \mathbf{u}; \theta^k), & \text{with probability }1-\varepsilon\\
        u\sim\operatorname{Uniform}(\mc U_d^k(\mathbf{p}_d^k)), & \text{with probability }\varepsilon
    \end{cases}
\end{equation}

Since~\eqref{eq:r} provides a natural way to provide feedback to the learning agent, the challenging part of establishing a multi-agent monitoring system boils down to designing informative traffic state representations that would allow learning good approximations of $Q(\mathbf{s},\mathbf{u})$. Moreover, the extracted state should be simple enough to accommodate real-time execution on the onboard hardware but rich enough to provide sufficient information about the state of the environment and the intentions of the rest of the fleet. 

For a a drone $d\in\mc D$ located at $\mathbf{p}_d^k$ at time $k\in\Z_T$, we propose a 13-element state vector $\mathbf{s}_d^k$. Let $\mathbf{p}_d^k$ correspond to the grid cell $(i,j)$ as specified in~\eqref{eq:1}. Then, the first seven elements of the state vector contain information about the drone's current position and the temporal importance within its field of view, scaled by the corresponding value of the idleness map. We will denote this set as $\tilde{\mc Z}_d^k$. The construction of the remaining six elements is detailed below.  

Let $\mc V_d^{\text{down}}$ denote the set of all non-obstacle cells $(i',j')$ in the idleness map that are in the $j$-th column but located below the cell $(i,j)$, i.e.,
\small$$\mc V_d^{\text{down}}=\{(i',j')\mid i\in\Z_{\mathbf{n}_x}, j\in\Z_{\mathbf{n}_y}, j'=j, i'>i, (i',j')\notin\mc O\}.$$\normalsize
If $|\mc V_d^{\text{down}}|=N_{d}^{\text{down}}$, then with a slight abuse of notation we can define $\nu_{d}^{\text{down}}\in\R$ as
$$\nu_{d}^{\text{down}}\defineas\frac{1}{N_{d}^{\text{down}}}\sum_{(i,j)\in\mc V_{d}^{\text{down}}}\mathbf{I}^k[i,j],$$
which represents the average idleness value of feasible cells located downward from the current position of the drone. Similarly, we can define $\nu_d^{\text{up}}$, $\nu_d^{\text{left}}$ and $\nu_d^{\text{right}}$ to describe the average idleness in the other possible movement directions. In addition to average idleness, each drone considers the position of the center of mass of the other drones to gain insight into the intentions of the rest of the swarm. Namely, we compute $i_{\text{cm}}\in\R$ and $j_{\text{cm}}\in\R$ as 
\begin{equation}
    i_{\text{cm}}=\frac{1}{N-1}\sum_{d'\in\mc D\setminus d}i_{d'}\:\land\:j_{\text{cm}}=\frac{1}{N-1}\sum_{d'\in\mc D\setminus d}j_{d'},
\end{equation}
where $(i_{d'},j_{d'})$ denotes the grid cell associated with the drone $d'\in\mc D\setminus d$, located at $\mathbf{p}_{d'}^k$ in line with~\eqref{eq:1}. The full state is now given by
\begin{equation}\label{eq:s}
    \mathbf{s}^k_d=\biggl\{\frac{i}{\mathbf{n}_x},\: \frac{j}{\mathbf{n}_y},\: \tilde{\mc Z}_d^k,\:\nu_d^{\text{down}},\:\nu_d^{\text{up}},\:\nu_d^{\text{left}},\:\nu_d^{\text{right}},\: \frac{i_{\text{cm}}}{\mathbf{n}_x},\: \frac{j_{\text{cm}}}{\mathbf{n}_y}   \biggr\}.
\end{equation}
With the definition of state $\mathbf{s}^k_d$ as in~\eqref{eq:s}, the corresponding reward for drone $d\in\mc D$ after taking the action $\mathbf{u}_d^k\in\mc U_{d}^k(\mathbf{p}_d^k)$ follows directly from~\eqref{eq:r}:
\begin{equation}
    \mathbf{r}_d^k(\mathbf{s}_d^k,\mathbf{u}_d^k,\mathbf{s}_d^{k+1})\defineas\alpha_{\textbf{T}}\overline{\mathbf{T}}_d(\mathbf{p}_{d}^{k+1}, \mathbf{p}_{d}^k)+\alpha_{\textbf{I}}\overline{\mathbf{I}}_d(\mathbf{p}_{d}^{k+1}, \mathbf{p}_{d}^k).
\end{equation}
With that in mind, in the following subsection we proceed to describe the learning procedure of the $\theta$ parameters.

\subsection{Pretraining of Q-function parameters}\label{sec:pret}

Since all drones are treated as equivalent, their individual experiences, recorded as tuples $b_k=(\mathbf{s}_d^k,\mathbf{u}_d^k,\mathbf{r}_d^k, \mathbf{s}_d^{k+1})$ while following different policies, can be combined to pretrain the parameters $\theta$ of a universal Q-function. Given that each drone has at most five possible actions, we model the Q-function as a neural network block that takes a 13-element state vector $\mathbf{s}_d^k$ as input and outputs a 5-element vector $q_d\in\R^5$, representing the Q values for each state-action pair.

Specifically, let $\mc B_d$ denote the memory buffer of drone $d\in\mc D$. At each epoch $m\in\Z_{N_{\text{ep}}}$, an experience batch $\overline{\mc B}\subset\bigcup_{d\in\mc D}\mc B_d$ of tuples $b_t=(\mathbf{s}^t,\mathbf{u}^t,\mathbf{r}^t, \mathbf{s}^{t+1})$ can be uniformly sampled and used to  update the parameters $\theta$. We perform $N_{\text{iter}}$ gradient descent steps in order to minimize the objective representing the discrepancy between the right and left-hand side of the Bellman equation:
\begin{equation}
    \mc L(\theta^m)=\frac{1}{|\overline{\mc B}|}\sum_{b_t\in\overline{\mc B}}(Y_t-Q(\mathbf{s}^t,\mathbf{u}^t;\theta^m)^2,
\end{equation}
where $Y_t\in\R$ is calculated as
\begin{equation}
    Y_t=\mathbf{r}^t+\gamma Q(\mathbf{s}^{t+1},\argmax_{\mathbf{u}'}Q(\mathbf{s}^{t+1},\mathbf{u}';\theta^m);\theta^m_{-}),
\end{equation}
in accordance with the Double DQN approach~\cite{10.5555/3016100.3016191}. Here, $Q(\cdot,\cdot;\theta^m_{-})$ represents a copy of the neural network representing the universal Q-function whose parameters we are trying to estimate. However, rather than training the second neural network, its parameters are hard-updated to $\theta^m$ every fixed number of epochs in order to improve the training stability and help avoid biased estimates of $Q$. 
\subsection{Online learning framework}
\begin {figure}
\centering
\begin{adjustbox}{max height=0.65\textwidth, max width=0.48\textwidth}
\begin{tikzpicture}[scale=1.0]
    \def\rows{4}
    \def\cols{10}
    \def\cellsize{0.5} 

    \foreach \x in {0,...,\cols} {
        \foreach \y in {0,...,\rows} {
            \draw (\x*\cellsize-2.5, 0-0.5) -- (\x*\cellsize-2.5, \rows*\cellsize-0.5);
            \draw (0-2.5, \y*\cellsize-0.5) -- (\cols*\cellsize-2.5, \y*\cellsize-0.5);
        }
    }
    \node[draw, rectangle, rounded corners=1mm, minimum width=10cm, minimum height=3cm, anchor=center,fill=orange!30,draw=orange!30] at (0,-4) {};

    \node (pic1) at (0.75-2.5, 0.25) {\includegraphics[height=.02\textwidth,width=.025\textwidth]{figures/drone.JPG}};
    \draw[-{Triangle[length=3pt, width=2.5pt]},line width=0.5pt, dashed](0.75-2.5, 0.25-1.25)--(0.75-2.5, 0.25);
    \node[ scale=0.8, black]() at (0.75-2.5-0.25, 0.25-1.25+0.25)  {$\mathbf{u}_1^k$};
    
    \node (pic2) at (2.25-2.5, 1.25) {\includegraphics[height=.02\textwidth,width=.025\textwidth]{figures/drone.JPG}};
    \draw[-{Triangle[length=3pt, width=2.5pt]},line width=0.5pt, dashed](2.25-2.5, 1.25-2.25)--(2.25-2.5, 1.25);
    \node[ scale=0.8, black]() at (2.25-2.5-0.25, 1.25-2.25+0.25)  {$\mathbf{u}_2^k$};
    
    \node (pic3) at (3.75-2.5, 0.75) {\includegraphics[height=.02\textwidth,width=.025\textwidth]{figures/drone.JPG}};
    \draw[-{Triangle[length=3pt, width=2.5pt]},line width=0.5pt, dashed](3.75-2.5, 0.75-1.75)--(3.75-2.5, 0.75);
    \node[ scale=0.8, black]() at (3.75-2.5-0.25, 0.75-1.75+0.25)  {$\mathbf{u}_3^k$};

    \node[ scale=0.8, black]() at (3.5, 0.75)  {$\{\mathbf{s}_d^k\}_{d\in\mc D}$};

    \node[draw, rectangle, minimum width=1cm, minimum height=1cm, anchor=center, fill=white] at (0.5,-4) {};
    \node[draw, rectangle, minimum width=1cm, minimum height=1cm, anchor=center, fill=white] at (-0.5,-4) {};

    \node[draw, rectangle, minimum width=1cm, minimum height=1cm, anchor=center, fill=white] at (3.5,-4) {};
    \node[draw, rectangle, minimum width=1cm, minimum height=1cm, anchor=center, fill=white] at (2.5,-4) {};

    \node[draw, rectangle, minimum width=1cm, minimum height=1cm, anchor=center, fill=white] at (-3.5,-4) {};
    \node[draw, rectangle, minimum width=1cm, minimum height=1cm, anchor=center, fill=white] at (-2.5,-4) {};

    \node (pic4) at (0.5,-4) {\includegraphics[height=.04\textwidth,width=.045\textwidth]{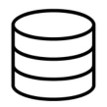}};
    \node[ scale=0.6, black]() at (-0.5,-4)  {$Q(\mathbf{s},\mathbf{u};\theta^k)$};

    \node (pic5) at (3.5,-4) {\includegraphics[height=.04\textwidth,width=.045\textwidth]{figures/buffer.jpg}};
    \node[ scale=0.6, black]() at (2.5,-4)  {$Q(\mathbf{s},\mathbf{u};\theta^k)$};

    \node (pic5) at (-2.5,-4) {\includegraphics[height=.04\textwidth,width=.045\textwidth]{figures/buffer.jpg}};
    \node[ scale=0.6, black]() at (-3.5,-4)  {$Q(\mathbf{s},\mathbf{u};\theta^k)$};    

    \node[draw, rectangle, rounded corners=1mm, minimum width=9cm, minimum height=1cm, anchor=center, fill=blue!10, draw=blue!10] at (0,-1.5) {};
    \node[ scale=0.8]() at (0.0, -1.5)  {\textbf{Central coordinator:} solves the optimization problem~\eqref{maxi:op2} };
    
    \node[draw, rectangle, rounded corners=1mm, minimum width=9cm, minimum height=1cm, anchor=center, fill=blue!10, draw=blue!10] at (0,-6.5) {};
    \node[ scale=0.8, black]() at (0.0, -6.5)  {\textbf{Central coordinator:} Updating a shared $Q(\mathbf{s},\mathbf{u};\theta)$ using $\overline{\mc B}\sim\bigcup_{d\in\mc D}\mc B_d$};
    
    \draw[-{Triangle[length=3pt, width=2.5pt]},line width=0.5pt](0.5, -4.5)--(0.5, -6);
    \draw[-{Triangle[length=3pt, width=2.5pt]},line width=0.5pt](3.5, -4.5)--(3.5, -6);
    \draw[-{Triangle[length=3pt, width=2.5pt]},line width=0.5pt](-2.5, -4.5)--(-2.5, -6);

    \draw[-{Triangle[length=3pt, width=2.5pt]},line width=0.5pt](-0.5, -6)--(-0.5, -4.5);
    \draw[-{Triangle[length=3pt, width=2.5pt]},line width=0.5pt](2.5, -6)--(2.5, -4.5);
    \draw[-{Triangle[length=3pt, width=2.5pt]},line width=0.5pt](-3.5, -6)--(-3.5, -4.5);

    \draw[-{Triangle[length=3pt, width=2.5pt]},line width=0.5pt](-0.5, -3.5)--(-0.5, -2);
    \node[ scale=0.8, black]() at (-0.75,-2.75)  {$q_2$};
    \draw[-{Triangle[length=3pt, width=2.5pt]},line width=0.5pt](2.5, -3.5)--(2.5, -2);
    \node[ scale=0.8, black]() at (-3.75,-2.75)  {$q_1$};
    \draw[-{Triangle[length=3pt, width=2.5pt]},line width=0.5pt](-3.5, -3.5)--(-3.5, -2);
    \node[ scale=0.8, black]() at (2.25,-2.75)  {$q_3$};

    \draw[-{Triangle[length=3pt, width=2.5pt]},line width=0.5pt](-2.75, 1.75)--(-1.75, 1.75);
    \node[ scale=0.8, black]() at (-1.75, 2)  {$y$};
    \draw[-{Triangle[length=3pt, width=2.5pt]},line width=0.5pt](-2.75, 1.75)--(-2.75, 0.75);
    \node[ scale=0.8, black]() at (-3, 0.75)  {$x$};

    \node[ scale=0.8, black]() at (-0.75,-5.25)  {$\theta^k$};
    \node[ scale=0.8, black]() at (-3.75,-5.25)  {$\theta^k$};
    \node[ scale=0.8, black]() at (2.25,-5.25)  {$\theta^k$};

    \node[ scale=0.8, black]() at (0.75,-5.25)  {$\mc B_2$};
    \node[ scale=0.8, black]() at (-2.25,-5.25)  {$\mc B_1$};
    \node[ scale=0.8, black]() at (3.75,-5.25)  {$\mc B_3$};
    
    \draw[-{Triangle[length=3pt, width=2.5pt]},line width=0.5pt](2.5, 0.5)--(5.5, 0.5)--(5.5,-4)--(5,-4);

    \draw[-{Triangle[length=3pt, width=2.5pt]},line width=0.5pt](-1.5, -3.25)--(-1.5,-4)--(-1,-4);
    \draw[-{Triangle[length=3pt, width=2.5pt]},line width=0.5pt](1.5, -3.25)--(1.5,-4)--(2,-4);
    \draw[-{Triangle[length=3pt, width=2.5pt]},line width=0.5pt](-4.5, -3.25)--(-4.5,-4)--(-4,-4);

    \node[ scale=0.8, black]() at (-1.5,-3.0)  {$\mathbf{s}_2^k$};
    \node[ scale=0.8, black]() at (-4.5,-3.0)  {$\mathbf{s}_1^k$};
    \node[ scale=0.8, black]() at (1.5,-3.0)  {$\mathbf{s}_3^k$};

    \node[ scale=0.8, black]() at (0.0,-3.25)  {Drone 2};
    \node[ scale=0.8, black]() at (-3.0,-3.25)  {Drone 1};
    \node[ scale=0.8, black]() at (3.0,-3.25)  {Drone 3};

    \node[ scale=0.8, black, rotate=90]() at (-5.25,-4.0)  {Swarm of drones $\mc D$};

\end{tikzpicture}
\end{adjustbox}
    \caption{The schematic illustrates the online learning framework for a three-drone swarm. When centralized coordination is active, each iteration samples a batch of data from the combined experiences of all drones to update the shared Q-function parameters which are then used to obtain the drone actions for the current time step.} 
\label{fig:online}
\end{figure}

Apart from learning the parameters of the Q-network from historical data, the proposed monitoring architecture also allows for online learning as illustrated in Figure~\ref{fig:online}. After loading the pretrained weights on each of the drones, at every time step $k\in\Z_T$, each drone first extracts a personal state vector from the traffic environment $\mathbf{s}_d^k$. Then, if the centralized coordination is enabled, the coordinator samples a batch $\overline{B}_k$ from $\bigcup_{d\in\mc D}\mc B_d$, uses it to perform one update epoch as described in Section~\ref{sec:pret}, and provides a copy of the updated parameter vector $\theta^k$ to each of the drones. Based on the observed states, the drones then compute individual vectors $q_d\in\R^5$ and return them to the central coordinator who solves the optimization problem~\eqref{maxi:op2} to determine the action of each drone. After executing their assigned actions $\mathbf{u}_d^k$, each drone updates its position and receives a patrolling reward $\mathbf{r}_d^k$, allowing them to store a transition tuple $(\mathbf{s}_d^k,\mathbf{u}_d^k,\mathbf{r}^k_d,\mathbf{s}_d^k)$ in their buffer $\mc B_d$ for future use. However, if central coordination fails, the system can continue operating, albeit with reduced performance. Specifically, if communication between drones is disrupted, they must rely on the last received copies of the $\theta$ parameters and only their locally updated idleness maps. The online learning process can proceed with the adjustment that, at each iteration, the training batch must be sampled solely from each drone’s individual data.

In the following section, we present two case studies demonstrating the performance of the RL-based approach in comparison to traditional path planning techniques. 
 
\section{Case study}\label{sec:example}

In this case study, we consider an urban map split into $20\times 30$ grid cells monitored by a swarm of 4 drones $\mc D=\{\triangle,\square,\circ,\star\}$. The observed time horizon is $T=2000$ steps and the parameters used to update the idleness map are given by $\eta=0.1$ and $\delta=0.025$. The shared Q-function is pretrained by executing $N_{\text{ep}}=30000$ training epochs, each containing $N_{\text{iter}}=30$ update iterations of the $\theta$ parameters. The size and the activation function of each fully-connected layer in the neural network representing the Q-function are given in Table~\ref{tab:1}, and the training batch size used was 32.
\begin{table}[t!]
    \renewcommand{\arraystretch}{1.3} 
    \centering
     \caption{Neural network structure representing the universal Q-function $Q(\mathbf{s},\mathbf{u};\theta)$.}\vspace{1ex}
    \label{tab:1}
    \begin{tabular}{c|c|c|c|c}
        
        \makecell{Input \\ layer} & \makecell{Hidden \\ layer 1} &  \makecell{Hidden \\ layer 2} &  \makecell{Hidden \\ layer 3} & \makecell{Output \\ layer} \\ \hline\hline
         $(13,2048)$ & $(2048,1024)$ & $(1024,256)$ & $(256,64)$ & $(64,5)$ \\ 
         ReLu & ReLu & ReLu & ReLu& Linear     \\ \hline
    \end{tabular}
\end{table}

We train the RL-patrolling agent in a completely artificial environment where the drone observations are governed by a priori chosen dynamic patterns. Specifically, we define a set of big slowly dissipating traffic disturbances $\mc S_{\text{big}}$ and a set of small periodic disturbances $\mc S_{\text{small}}$ that both contribute to the temporal importance observed by a drone. Here, big disturbances are used to model traffic phenomena such as accidents, traffic jams, etc., whereas the small ones are used to model the effects of traffic lights, traffic signs, etc. In particular, we assume that the observation when located at position $\mathbf{p}_d^k$, associated with cell $(i,j)$, is given by 
\begin{equation}
    z(\mathbf{p}_d^k)=\sum_{p\in\mc S_{\text{big}}\cup\mc S_{\text{small}} }A_s(k)e^{-\frac{1}{2}[(i-i_s)^2+(j-j_s)^2]},
\end{equation}
where $(i_s,j_s)$ represents the cell containing the origin of a traffic disturbance and the amplitude $A_s(k)$ is given by
$$A_s(k)=\begin{cases}
    e^{-\frac{k}{\beta_1 T}}, & \text{if }s\in S_{\text{big}}\\
    \max(0;\:\sin(2\frac{\beta_2\pi}{T}\cdot k)), & \text{if }s\in\mc S_{\text{small}}
\end{cases},$$
for some $\beta_1,\beta_2>0$. To gather data needed for pretraining the neural network model, we collected tuples by running a swarm of randomly moving drones on a map featuring 4 large and 3 small disturbances. In contrast, the patrolling algorithms were tested on a map with 2 large and 3 small disturbances, different from the ones used during training. 

After analyzing and comparing the performance of the RL-based agent with more traditional benchmarks in the artificial environment, in Section~\ref{sec:cs2}, we directly transfer the trained model to a setup based on real traffic data from the city of Shenzhen in China. In contrast to the training environment, the observation model here is completely stochastic and reflects dynamic traffic changes appearing during rush hours.    

\subsection{Swarm model comparison}

\begin {figure}
\centering
\begin{adjustbox}{max height=0.55\textwidth, max width=0.48\textwidth}
\begin{tikzpicture}[scale=1.0]

    \node (pic1) at (-2.2, 0.0) {\includegraphics[height=.18\textwidth,width=.25\textwidth]{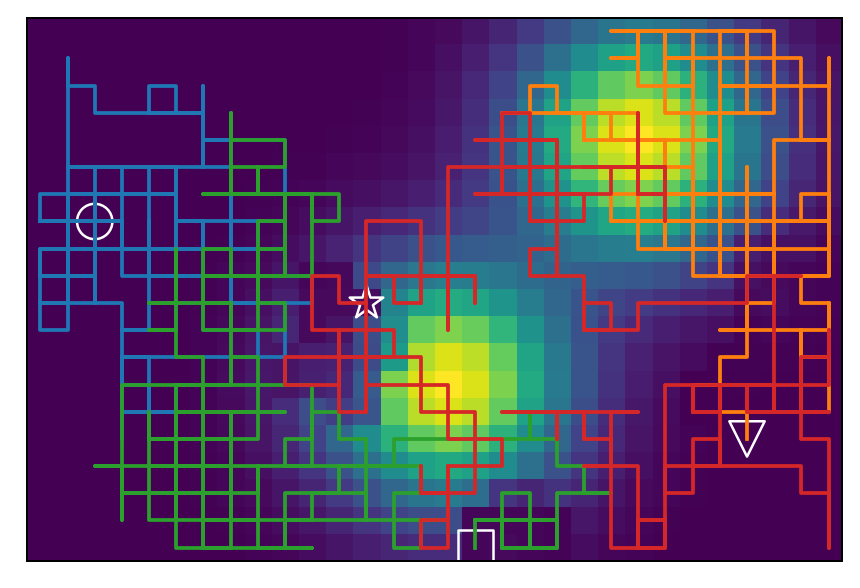}};

    \node (pic2) at (2.2, 0.0) {\includegraphics[height=.18\textwidth,width=.25\textwidth]{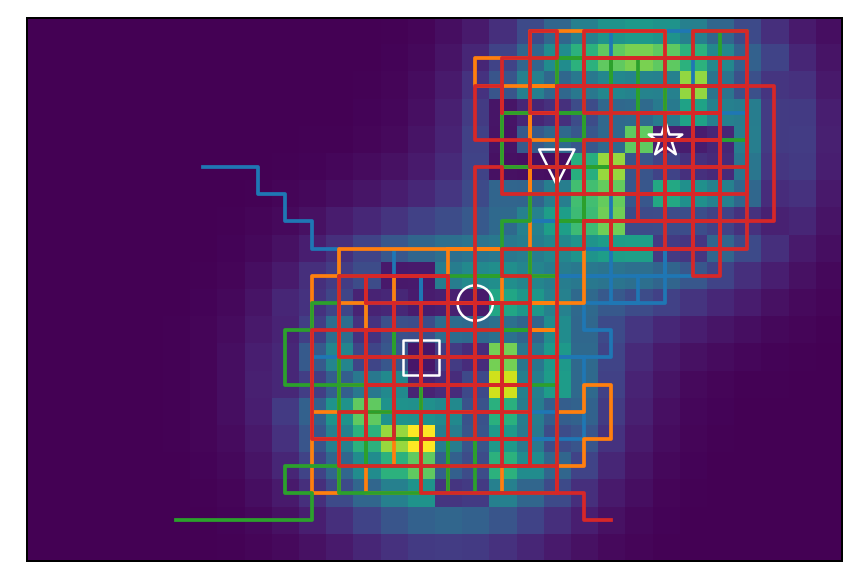}};

    \node (pic3) at (-2.2, -4.0) {\includegraphics[height=.18\textwidth,width=.25\textwidth]{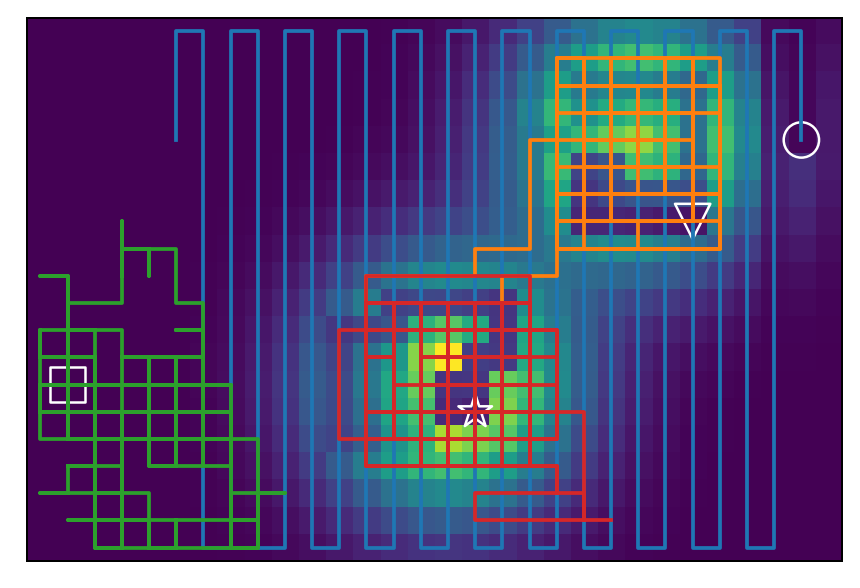}};

    \node (pic4) at (2.2, -4.0) {\includegraphics[height=.18\textwidth,width=.25\textwidth]{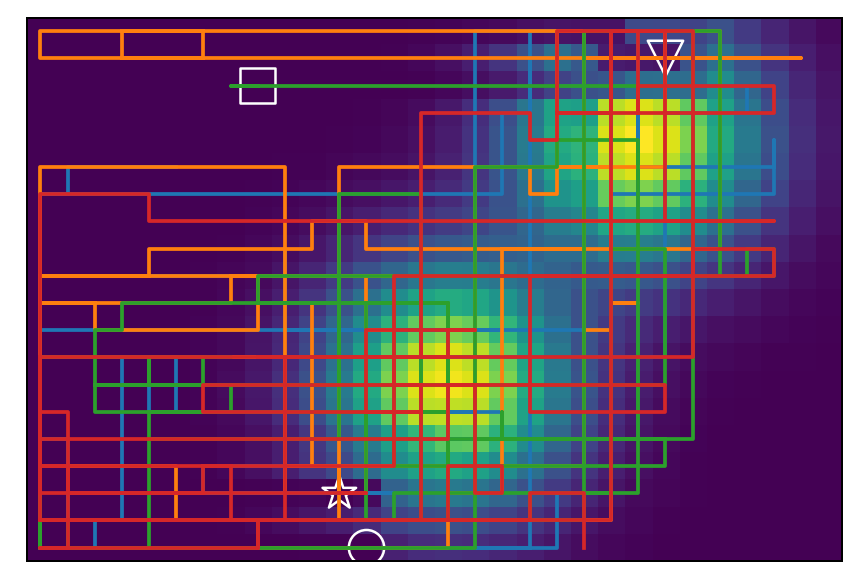}};
    
    \node[ text=black, scale=1.0](t1) at (-2.2, -2.0){(a) Random};

    \node[ text=black, scale=1.0](t2) at (2.2, -2.0){(b) Greedy};

    \node[ text=black, scale=1.0](t3) at (-2.2, -6.1){(c) Sweeping};

    \node[ text=black, scale=1.0](t4) at (2.2, -6.1){(d) RL based};
\end{tikzpicture}
\end{adjustbox}
    \caption{Drone trajectories for various swarm types are shown, with cell colors representing the value $\mathbf{T}^k(\mathbf{p})\mathbf{I}^k(\mathbf{p})$. Brighter cells indicate areas of high unvisited importance, while darker cells correspond to recently visited or less important regions.} \vspace{-1.5em}
\label{fig:comp}
\end{figure}
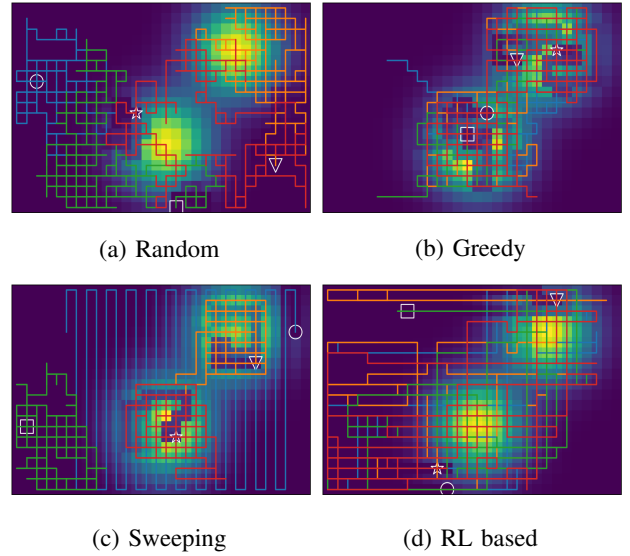

Apart from the `\textbf{RL-based}' and the `\textbf{random}' swarms of drones, we also test the `\textbf{greedy}' and the `\textbf{sweeping}' ones of the same size. In a greedy swarm, each drone chooses the action $\mathbf{u}_d^k$ that yields $\mathbf{p}_d^{k+1}=\overline{\mathbf{p}}$, where $\overline{\mathbf{p}}$ satisfies:
$$\overline{\mathbf{p}}\in\argmax_{\mathbf{p}\in\operatorname{Adj}(\mathbf{p}_d^k)}\mathbf{I}^k(\mathbf{p})\mathbf{T}^k(\mathbf{p}).$$
The sweeping swarm consists of one drone following a zig-zag path that covers the entire map, along with two greedy drones and one random drone. For $\beta_1=0.7$ and $\beta_2=5$, Figure~\ref{fig:comp} shows the color-coded patrolling trajectories for each drone across different types of swarms. 

\begin{figure}
    \centering
    \begin{adjustbox}{ max width=0.55\textwidth}
    \input{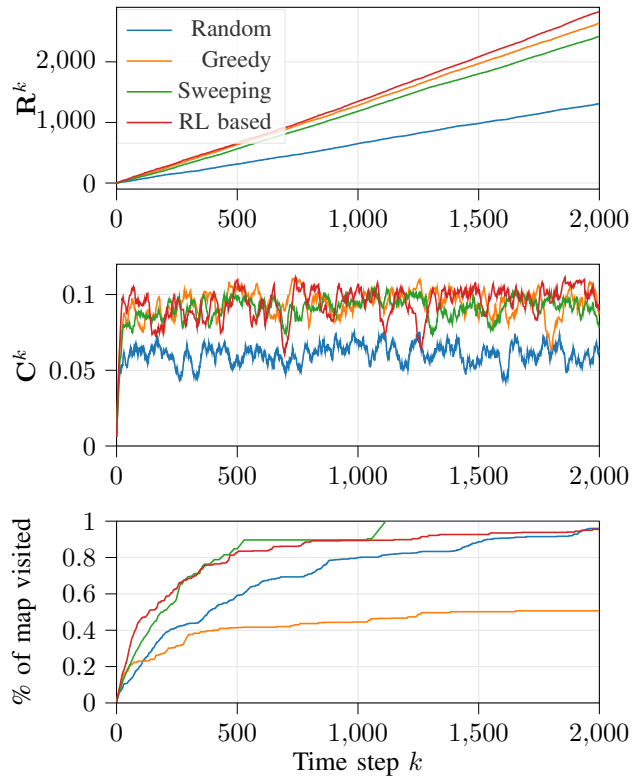}
    \end{adjustbox}
    \vspace{-0.5em}
    \caption{Performance comparison of different swarm models. For every $k\in\Z_T$, the first plot shows cumulative patrolling score obtained until that moment. The second one presents the evolution of the patrolling score over time, whereas the final one shows the percentage of cells visited from the beginning of simulation.}\vspace{-1.5em}
    \label{fig:score}
\end{figure}

The greedy swarm effectively converges to high-importance areas but fails to monitor regions along the boundaries. In contrast, the random swarm covers a larger portion of the map but lacks a systematic approach to revisiting high-importance regions. The sweeping and the RL-based ones seem to have established a good balance between exploring the unobserved areas and revisiting the high-importance regions. To assess performance quality, we plot the temporal evolution of the cumulative patrolling score $\mathbf{R}^k$, coverage score $\mathbf{C}^k$, and the percentage of the map visited over the time horizon $T$ in Figure~\ref{fig:score}. These visualizations are further supported by numerical data in Table~\ref{tab:2}. The results indicate that the RL-based agent achieves the highest cumulative patrolling score, with $\mathbf{R}_T^{\text{rl}}\approx 2830$, surpassing other methods. The greedy and sweeping agents also perform well, with scores of $\mathbf{R}_T^{\text{gr}}\approx 2630$ and $\mathbf{R}_T^{\text{sw}}\approx 2400$, respectively, while the random agent underperforms significantly with $\mathbf{R}_T^{\text{rd}}\approx 1300$. Furthermore, as indicated in Table~\ref{tab:2}, the RL-based agent’s average patrolling score over the entire time horizon, $\hat{\mathbf{R}}_T$, is 10.5\% higher than that of the second-best method, achieved by the greedy agent. As expected, in terms of percentage of the map visited during the horizon $T$, the sweeping agent performs best and visits all cells, while the greedy one performs significantly worse than all other approaches, visiting only about $51\%$ of the map. The coverage score, on the other hand, reveals that while the random agent visits most of the cells on the map, the time between two visits to the same cell is, on average, longer compared to other methods. Most of the time, the sweeping agent is slightly outperformed by both the RL-based and greedy agents, which is reflected in its lower average coverage score, $\hat{\mathbf{C}}^{\text{sw}}_T=0.090$, compared to the RL-based and greedy agents. Although the average coverage scores of the RL-based ($\hat{\mathbf{C}}^{\text{rl}}_T = 0.094$) and greedy ($\hat{\mathbf{C}}^{\text{gr}}_T = 0.093$) agents are similar, the RL-based agent outperforms the others in 46.56\% of the time horizon, achieving the highest value of $\mathbf{C}^k$ among all four agents. This gives the RL-based agent a slight advantage overall in this regard. 
\begin{table}[!t]
\begin{center}
 \renewcommand{\arraystretch}{1.3}
 \caption{Attained profits for different planning horizons}\vspace{1ex}
 \label{tab:2}
 \begin{tabular}{c|c|c|c|c}
 Swarm  & $\hat{\mathbf{R}}_T$ & $\hat{\mathbf{C}}_T$ & \makecell{$\%$ of time \\ with max $\mathbf{C}^k$} & \makecell{$\%$ of map \\ covered} \\ 
 \hline\hline
 Random   &  0.649 & 0.060  & 0.05  & 96.0 \\
 Greedy   &  1.299 & 0.093  & 37.37 & 50.7  \\
 Sweeping &  1.188 & 0.090  & 16.01 & 100.0 \\
 RL-based &  1.373 & 0.094  & 46.56 & 95.5 \\
\hline 
\end{tabular}
\end{center}
\end{table}

\subsection{Case study based on real traffic data}\label{sec:cs2}

We demonstrate that the design of our RL-based swarm enables the direct transfer of models learned in a fully artificial environment to a case study based on real traffic data, highlighting the potential for real-world application.

\begin {figure*}
\centering
\begin{adjustbox}{max height=0.65\textwidth, max width=\textwidth}

\begin{tikzpicture}[scale=1.0]

    \node (pic1) at (-2.2, 0.0) {\includegraphics[height=.18\textwidth,width=.25\textwidth]{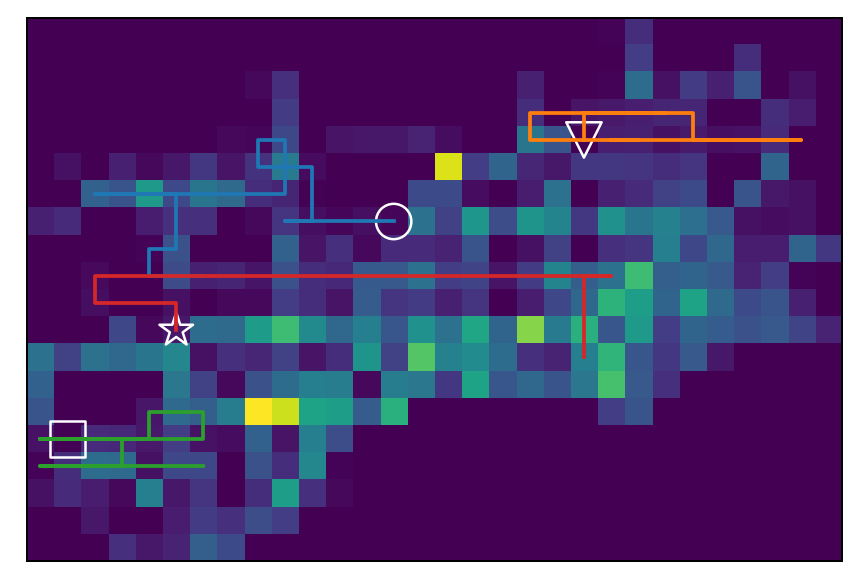}};

    \node (pic2) at (2.2, 0.0) {\includegraphics[height=.18\textwidth,width=.25\textwidth]{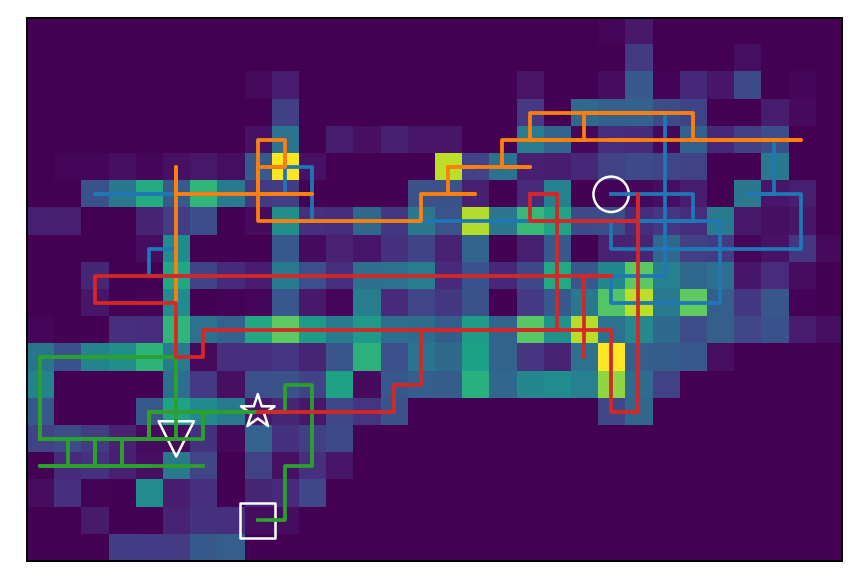}};
    
    \node (pic3) at (-6.6, 0.0) {\includegraphics[height=.18\textwidth,width=.25\textwidth]{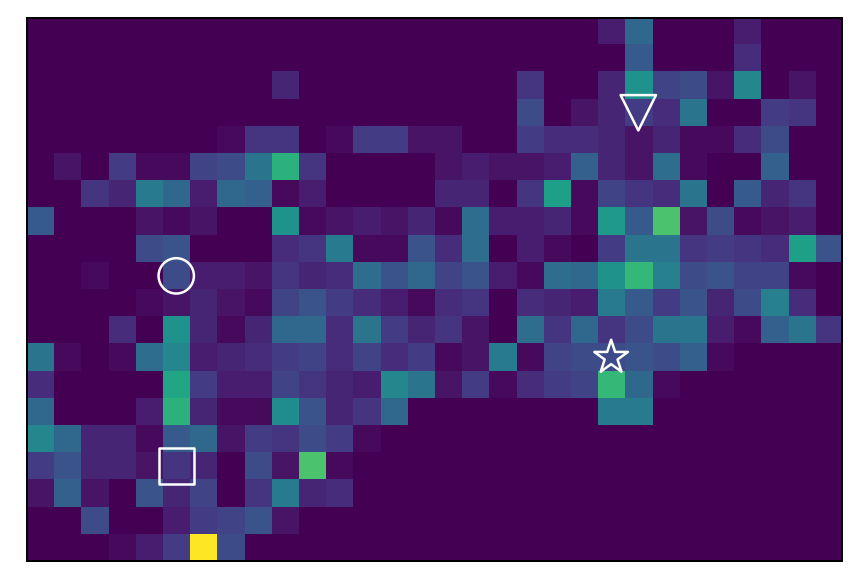}};

    \node (pic4) at (6.6, 0.0) {\includegraphics[height=.18\textwidth,width=.25\textwidth]{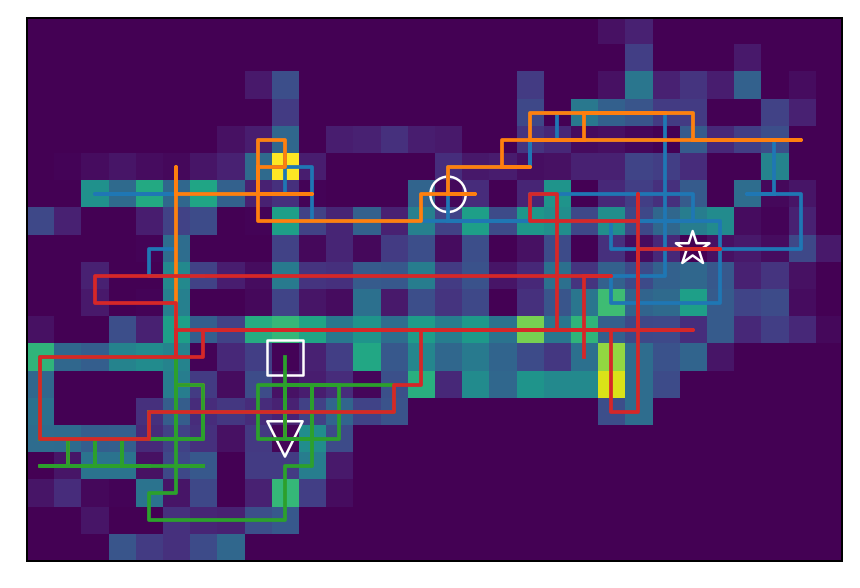}};
    
    \node[ text=black, scale=1.0](t1) at (-2.2, -2.1){(b) $k=40$};

    \node[ text=black, scale=1.0](t2) at (2.2, -2.1){(c) $k=90$};

    \node[ text=black, scale=1.0](t3) at (-6.6, -2.1){(a) $k=0$};

    \node[ text=black, scale=1.0](t4) at (6.6, -2.1){(d) $k=140$};
    
\end{tikzpicture}
\end{adjustbox}
\caption{Temporal evolution of drone trajectories in an RL-based swarm, demonstrated through a case study using real traffic data from the city of Shenzhen.}
\label{fig:real}
\end{figure*}
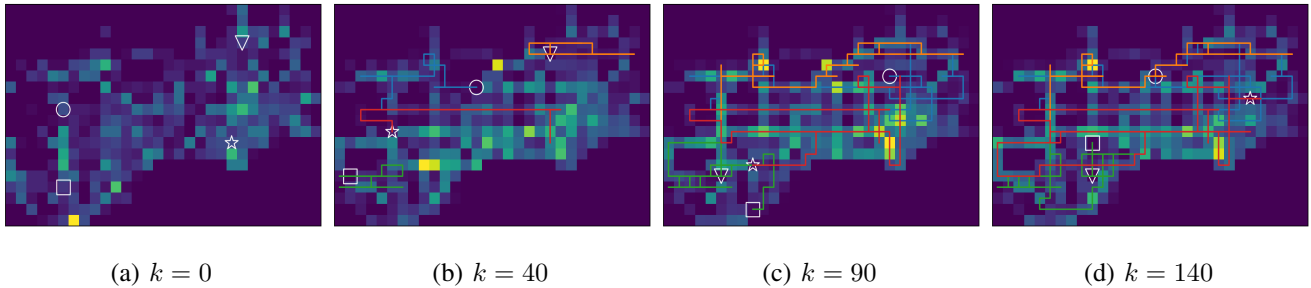

We address the problem of monitoring road occupancy during rush hour periods in the city of Shenzhen in China. Similar to the illustration in Figure~\ref{fig:setup}, we assume the city area is divided into a $20\times 30$ grid where drones can move between cells and count the number of vehicles in each of them. To deploy the RL-based agent, the observed vehicle counts must be converted into a temporal importance value that reflects traffic conditions. This transformation is done using an experimentally derived Macroscopic Fundamental Diagram~\cite{GEROLIMINIS2008759}, which maps vehicle counts to expected speeds at the cell level. By incorporating the road speed limits for each cell, we can then establish a temporal importance mapping that aligns with Definition~\ref{def:2}. We simulate a time interval of $T_{\text{sim}}=3 \text{ h}$, representing the spatio-temporal evolution of traffic conditions during one of the rush hour periods of the day. In this period, the traffic volume, which includes both private and ride-hailing vehicles, increases significantly, potentially causing congestion in high-demand areas that require monitoring. Taking into account the real dimensions of the map in Figure~\ref{fig:setup}, we estimate that the distance between the centers of adjacent grid cells is approximately $500 \text{ m}$. Given the average drone speed of $v_{\text{drone}} = 24 \text{ km/h}$, it takes about $\Delta t = 1.25 \text{ min}$ for a drone to traverse from the center of one cell to the next. This results in a monitoring horizon of $T = 144$ steps over the simulation period of three hours.
Figure~\ref{fig:real} shows the temporal evolution of drone trajectories in an RL-based swarm. As previously stated, apart from the monitoring horizon, no other parameter of the learned Q-function was changed compared to the training phase. In contrast to the training environment, the observation distribution in this case is governed by stochastic demand patterns, making it more challenging to monitor. As illustrated in Figure~\ref{fig:real}, drones successfully navigate in a fully autonomous way, demonstrating potential for `sim-to-real' training in more complex environments. 
\section{Conclusions}\label{sec:conclusion}

In this paper, we presented a universal end-to-end solution for monitoring any traffic phenomena using a swarm of drones in urban environments. The proposed approach is fully scalable with respect to the size of the swarm as a single Q function is trained to govern the action-selection process for each drone using the Double DQN procedure.  Compared to traditional patrolling methods, our approach outperforms in terms of both patrolling efficiency and coverage scores, all while being trained in a simple, entirely artificial environment. Furthermore, the proposed framework is resilient to potential failures of a central coordinator, as it can also run in a fully distributed online regime if necessary, albeit with a potential reduction of performance quality.     

Given the extensive literature on recurrent models, such as LSTMs, that are capable of reasoning about sequential data, it would be valuable in the future to explore whether adapting these models could further improve patrolling performance. From a practical standpoint, it would also be useful to test the scalability of the model, specifically examining the largest model size that could be deployed on low-cost monitoring hardware. Finally, since the proposed patrolling planner does not account for changes in drone altitude, it would be interesting to integrate it with a lower-level obstacle avoidance controller and test its effectiveness in patrolling missions involving ground vehicles.   

\bibliographystyle{IEEEtran}
\bibliography{references.bib}
\end{document}